\documentclass{sigchi}

\CopyrightYear{2018} 
\setcopyright{acmcopyright} 
\conferenceinfo{CHI 2018,}{April 21--26, 2018, Montreal, QC, Canada}
\isbn{978-1-4503-5620-6/18/04}\acmPrice{$15.00}
\doi{https://doi.org/10.1145/3173574.3173989}

\toappear{To appear at CHI 2018}

\usepackage{balance}       
\usepackage{graphics}      
\usepackage[T1]{fontenc}   
\usepackage{txfonts}
\usepackage{mathptmx}
\usepackage[pdflang={en-US},pdftex]{hyperref}

\usepackage[usenames, dvipsnames]{color}

\usepackage{booktabs}
\usepackage{textcomp}

\usepackage{microtype}        
\usepackage{ccicons}          

\usepackage{todonotes}

\usepackage{tabularx,ragged2e}
\newcolumntype{C}{>{\Centering\arraybackslash}X}

\def\plaintitle{SIGCHI Conference Proceedings Format}

\def\emptyauthor{}
\def\plainkeywords{Authors' choice; of terms; separated; by
  semicolons; include commas, within terms only; required.}

\makeatletter
\def\url@leostyle{%
  \@ifundefined{selectfont}{
    \def\UrlFont{\sf}
  }{
    \def\UrlFont{\small\bf\ttfamily}
  }}
\makeatother
\def\pprw{8.5in}
\def\pprh{11in}

\definecolor{linkColor}{RGB}{6,125,233}
\hypersetup{%
  pdftitle={\plaintitle},
  pdfauthor={\emptyauthor},
  pdfkeywords={\plainkeywords},
  pdfdisplaydoctitle=true, 
  bookmarksnumbered,
  pdfstartview={FitH},
  colorlinks,
  citecolor=black,
  filecolor=black,
  linkcolor=black,
  urlcolor=linkColor,
  breaklinks=true,
  hypertexnames=false
}

\begin{document}

\title{Touch Your Heart: A Tone-aware Chatbot for Customer Care on Social Media}


\numberofauthors{1}

\author{%
	  \alignauthor Tianran Hu$^{*}$, Anbang Xu$^{\dag}$, Zhe Liu$^{\dag}$, Quanzeng You$^{*}$, Yufan Guo$^{\dag}$, Vibha Sinha$^{\dag}$,\\ Jiebo Luo$^{*}$, Rama Akkiraju$^{\dag}$\\
	  \affaddr{
	  	$^*$University of Rochester
	  	$^\dag$IBM Research - Almaden
	  }\\
	  \email{
	  \{thu,qyou,jluo\}@cs.rochester.edu,\\ \{anbangxu, liuzh, guoy, vibha.sinha, akkiraju\}@us.ibm.com
	  }
}

\maketitle

\begin{abstract}

Chatbot has become an important solution to rapidly increasing customer care demands on social media in recent years. However, current work on chatbot for customer care ignores a key to impact user experience - tones. In this work, we create a novel tone-aware chatbot that generates toned responses to user requests on social media. We first conduct a formative research, in which the effects of tones are studied. Significant and various influences of different tones on user experience are uncovered in the study. With the knowledge of effects of tones, we design a deep learning based chatbot that takes tone information into account. We train our system on over 1.5 million real customer care conversations collected from Twitter. The evaluation reveals that our tone-aware chatbot generates as appropriate responses to user requests as human agents. More importantly, our chatbot is perceived to be even more empathetic than human agents.

\end{abstract}

\category{H.5.3.}{Information Interfaces and Presentation
  (e.g. HCI)}{Group and Organization Interfaces}{}{}


\section{Introduction}

Chatbot for customer care on social media has drawn much attention from both industry~\cite{ibm,micro,facebook} and academia~\cite{do2016empathic,candello2017typefaces} in recent years. As the wide adoption of social media, more and more users are seeking service on the new platform. It is reported that, in general, 23\% U.S. customers have used at least one company's social media site for servicing~\cite{mitchell2015state}, and the percentage increases to 67\% for online shopping customers~\cite{bworld}. However, due to the massive volume of user requests on social media, manual customer care often fails users' expectations. For example, the average response time to user requests on social media is 6.5 hours, which is a lot longer than users' expected waiting time of one hour, and many requests are even not responded at all~\cite{waitingtime}. Therefore, chatbot systems that respond to user requests automatically become a nature solution to improve user experience.

\begin{figure}
\centering
\includegraphics[width=1 \columnwidth]{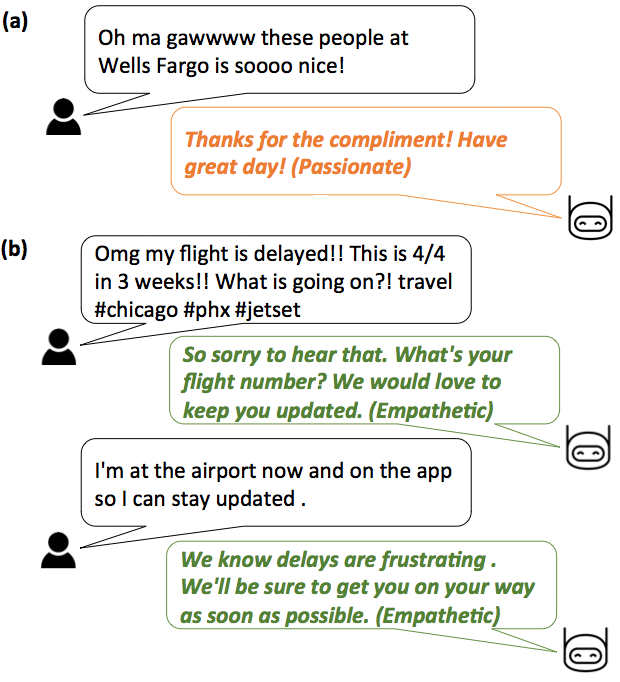}
\caption{Two examples of the responses generated by our tone-aware chatbot. (a) demonstrates a one-round conversation between a user and the chatbot in a passionate tone (in \textcolor{YellowOrange}{orange}). (b) demonstrates a two-round conversation, in which the user and chatbot responses to each other alternatively. In the conversation, the chatbot addresses user requests in a empathetic tone (in \textcolor{OliveGreen}{green}).}~\label{fig:conv_exp}
\vspace{-2em}
\end{figure}

There has been a long history of chatbot powered by various techniques~\cite{mairesse2007personage,lasecki2013chorus}. Currently, deep learning based techniques have the state-of-the-art performance, and significantly outperform transitional rule-based models~\cite{sutskever2014sequence}. In terms of customer care, deep learning based chatbot reportedly generates proper responses to user requests~\cite{xu2017new}, showing its encouraging application perspective. However, previous work on chatbot for customer care usually only focuses on generating grammatically correct responses, ignoring other factors that could affect user experience. In this paper, we propose a novel tone-aware chatbot. The proposed chatbost is inspired by very recent deep learning techniques~\cite{li2016persona,zhou2017emotional}, and takes an important factor of customer care -- tones -- into account. Figure~\ref{fig:conv_exp} shows two example conversations between users and our tone-aware chatbot.

Much work suggests that tones used in responses to users are key to a satisfactory service~\cite{davidow1998effects,morris1988many}. For example, courtesy tone has a significant effect on outcome satisfaction~\cite{zhang2011effects}, and empathetic tone reduces user stress and results in more engagement~\cite{brave2005computers}. However, in the context of social media customer care, the effects of tones are not yet systematically studied. Therefore, we first conduct a formative study to gain the knowledge. We identify eight typical tones in customer care. The identified tones include \textbf{\textit{anxious}}, \textbf{\textit{frustrated}}, \textbf{\textit{impolite}}, \textbf{\textit{passionate}}, \textbf{\textit{polite}},  \textbf{\textit{sad}}, \textbf{\textit{satisfied}}, and \textbf{\textit{empathetic}}. Regression analysis is then performed to study the effects of the selected tones. The results uncover significant and various impacts of different tones. For example, it is observed that empathetic tone significantly reduces users' negative emotion, such as frustration and sadness. Also, passionate tone cheers users up, and increases service satisfaction. According to the analysis results, we are able to identify the tones that are beneficial for customer care. Furthermore, we study the representative words of these beneficial tones. Our chatbot integrates the learned knowledge from the formative study, and targets at generating responses to user requests in these tones.

We evaluate our system from two aspects: 1) the response quality, i.e. if the system generates proper responses to user requests, and 2) the intensities of embedded tones, i.e. if human could perceive tones embedded in the generated responses. Therefore, appropriateness, helpfulness, and tone intensities are annotated for the responses by both out chatbot and real human agents. Statistical tests are conducted on the annotation data to compare our chatbot and human agents. The test results indicate that our chatbot can perform as appropriately as human agents. Meanwhile, it is observed that annotators can correctly perceive the tones embedded in the responses. More importantly, our chatbot is perceived to respond even more empathetically than human agents.


Our main contributions are three-fold:

1) We systematically study the effects of tones in the context of social media customer care. The results indicate the significant and various impacts of different tones on user experience. The study sheds light on a better understanding on how tones affect service quality of social media customer care.

2) We propose a novel chatbot system that not only generates proper responses to user requests, but also embeds tones in the responses. To the best of our knowledge, this is the first work considering tones for customer care chatbot on social media. 

3) Our system is validated by human judgments. The evaluation reveals that our system generates both proper and tone-aware responses to user requests. More importantly, the responses generated by our system is perceived even more empathetic than the responses by human agents.

\begin{table*}
{\renewcommand\arraystretch{1.25}
\begin{tabular}{|l|l|l|l|l|} \hline
Tones & \multicolumn{2}{l|}{Definition} & \multicolumn{2}{l|}{Example Tweets} \\ 
\hline
\hline

\textbf{\textit{Empathetic}} &  \multicolumn{2}{p{5cm}|}{\raggedright An affective mode of understanding that involves emotional resonance.} &  \multicolumn{2}{p{10cm}|}{\raggedright \textit{``Hi, sorry to see you are having trouble. Was 2 payments taken for ebay fees or have you sold something? Thanks''}}\\
\hline

\textbf{\textit{Passionate}} &  \multicolumn{2}{p{5cm}|}{\raggedright Showing enthusiasm and interest.} & \multicolumn{2}{p{10cm}|}{\raggedright  \textit{``We're excited to see you on board soon too! :-D ''}} \\
\hline

\textbf{\textit{Satisfied}} & \multicolumn{2}{p{5cm}|}{\raggedright An affective response to perceived user experience.} & \multicolumn{2}{p{10cm}|}{\raggedright  \textit{``I got my replacement for the damaged laptop . I just dropped by to tell you I LOVE AMAZON EVEN MORE <3''}}\\
\hline

\textbf{\textit{Polite}} & \multicolumn{2}{p{5cm}|}{\raggedright Being rational and goal-oriented.} & \multicolumn{2}{p{10cm}|}{\raggedright  \textit{``Could you please help me out in this matter? Thank you!''}}\\
\hline

\textbf{\textit{Impolite}} & \multicolumn{2}{p{5cm}|}{\raggedright Being disrespectful and rude.} & \multicolumn{2}{p{10cm}|}{\raggedright  \textit{``@tmobile is a piece of shit for a company.''}}\\
\hline

\textbf{\textit{Sad}} & \multicolumn{2}{p{5cm}|}{\raggedright Being unpleasant and passive.} & \multicolumn{2}{p{10cm}|}{\raggedright  \textit{``Netflix isn't working and I want to cry.''}}\\
\hline

\textbf{\textit{Frustrated}} & \multicolumn{2}{p{5cm}|}{\raggedright Feeling annoyed and irritable.} & \multicolumn{2}{p{10cm}|}{\raggedright  \textit{``@AmazonHelp u pathetic people, I still dont have my package, and no one is committing when it will get delivered.''}}\\
\hline

\textbf{\textit{Anxious}} & \multicolumn{2}{p{5cm}|}{\raggedright Experiencing worry, unease, or nervousness. } & \multicolumn{2}{p{10cm}|}{\raggedright  \textit{``Checked the tracking info for my package. No update in almost two days! it's stuck somewhere??? @UPSHelp''} }\\
\hline

\end{tabular}}

\caption{The definitions and example tweets of eight major tones identified from real customer care conversations. Please note that since the examples tweets are collected from real data, the texts may contain swear words and nonstandard writings. }~\label{tab:tone_def}
\vspace{-2em}
\end{table*}

\section{Related Work}
\subsection{Studies on Customer Care}

There has been a lot of work on customer care quality, and the factors that could affect user experience. For example, one factors is the speed with which agents respond to users~\cite{blodgett1997effects, clark1992consumer}. Although the response time is important intuitively, research results indicate that it is not always true. Boshoff reported that time is not a dominate factor until the waiting time becomes too long~\cite{boshoff1997experimental}, and Morris reported that a quick response does not make user more satisfied~\cite{morris1988many}. Compensation is another well studied factor of customer care. Different from the time factor, this factor is reportedly of significant effect on user satisfaction~\cite{mccollough2000effect}. Furthermore, compensation could also increase customer repurchase intention~\cite{sparks2001justice} and word-of-mouth activities~\cite{mack2000perceptions}. Other factors studied include company policies for customer care~\cite{nyer2000investigation, davidow1998effects}, conversation skills of agents~\cite{davidow2000bottom}, and so on. The effects of tones for customer care are also emphasized in literature, we review the related work in the next section. ~\cite{davidow2003organizational} gives a nice survey of studies on the factors that affect user experience.

Since customer care on social media is gaining increasing popularity in recent years, more and more work has focused on the new platform~\cite{oraby2017may, herzig2016classifying}. Laroche et al. pointed out that brand activities on social media could fasten the relationship with customer, and further increase customer loyalty~\cite{laroche2013or}. Istanbulluoglu studied the time factors on social media~\cite{istanbulluoglu2017complaint}. The author concluded that users expect shorter response time on social media than transitional channels. Also, the expected response time is different across platforms, for example, users expect to be responded more quickly on Twitter than on Facebook. Einwiller et al. studied the frequent actions taken by agents on social media~\cite{einwiller2015handling}. The most frequent action is inquiring further information, followed by expressing gratitude and regretting. It is reported that compensations are rarely offered on social media. The effects of conversation skills of agents on social media are discussed in~\cite{kang2013impact}. The authors reported that agents being conversational and friendly could increase the likelihood of purchase.

\subsection{Tones in Customer Care}

Much research~\cite{hasegawa2013predicting, li2015sentence} indicates that tones embedded in conversations have significant emotional impacts on conversation participants. Therefore, the application of tones in customer care has naturally drawn much attention. Morris pointed out that agents should not only offer help to users, but also address users' feelings, therefore, tones needed to be used in the responses~\cite{morris1988many}. Much previous work reveals the general benefit of applying tones in agent responses, such as increasing user attitude toward the company~\cite{martin1985consumer}, positive effect on word-of-mouth activity~\cite{tarp1982measuring}, and increasing user satisfaction~\cite{mccollough2000effect}. There is also work on the effects of specific tones. For example, Zhang et al. showed that apologetic tone has a positive effect on user satisfaction~\cite{zhang2011effects}. Meanwhile, agents being cheerful in their language could increase user positive emotions~\cite{herzig2016classifying}. Also, empathetic tone reportedly leads to more customer trust~\cite{brave2005computers}, and reduces customer stress~\cite{prendinger2005empathic}. In this paper, we systematically study the effects of different tones in customer care, and use the knowledge as guide for automatically generating toned responses to user requests. 

\subsection{Chatbot Systems}

There has been a long history of chatbot powered by various techniques. The early work on the topic is mainly rule based~\cite{ritter2011data} or retrieval based~\cite{yu2015ticktock}. However, these techniques are usually limited to small-scale data or close application domains. A possible solution to conducting free form conversations is Crowdsourcing chatbot system~\cite{chang2017evorus,huang2016there}. However, this type of systems is mainly based on human manual operations, therefore, not applicable to our task. 

Recently, thanks to the work on deep learning, large-scale and open-domain conversation generation has been investigated~\cite{li2016deep}. Currently, the sequence to sequence (seq2seq) models based on recurrent neural work have the state-of-the-art performance on the task~\cite{sutskever2014sequence, ha2016hypernetworks}. In practice, Xu et al.~\cite{xu2017new} applied the model on a chatbot system for customer care, and reported satisfactory performance on various metrics. However, the standard seq2seq model does not take meta information of data, such as tones in our case, into account. As a consequence, the model is not capable of controlling the styles of the output conversation. To overcome the limit, Li et al. reported a modified seq2seq model~\cite{li2016persona}. In the work, the authors controlled the output conversation by adding an indicator vector in the model, and generated sentences matching certain persona. Similarly, Zhou et al. proposed an automatic conversational system that generates conversations of various emotions~\cite{zhou2017emotional}. Inspired by the work, we propose a novel seq2seq model that is capable of controlling the tones of generated conversations.

\section{Data Collection and Preprocessing}

We collect our dataset from Twitter, a widely adopted social media platform for customer care. We first select 62 brands across various industries, such as Technology (e.g. Apple, HP), Airline (e.g. JetBlue, Southwest), Retail (e.g. Walmart, Target), and so on. We manually locate the customer care accounts of these brands on Twitter, for example @AppleSupport for Apple, and @JetBlue for JetBlue. We then follow the processes suggested in~\cite{kim2012you} to recover the conversations between brands' Twitter accounts (agents) and users. Specifically, we first download all the tweets sent by the brand accounts using Twitter Developer API\footnote{https://dev.twitter.com/} from Aug $1^{st}$ 2016 to Jun $1^{st}$ 2017. The API returns many types of information of a tweet, including if the tweet is a reply to another tweet, and the ID of the replied tweet (Reply\_ID). Note that, each tweet is assigned an unique ID. We then download the tweets that agents responded to using their Reply\_IDs, and check if these tweets are also responses to other tweets. We keep tracing the conversation chains in such a bottom up manner, until the tweets we download are the initial utterances of the conversations (the initial tweets' Reply\_ID are None). By doing this, we collect totally 3.5 million tweets sent by either users or brand accounts.

By matching the ID and Reply\_ID of the collected tweets, we restore the conversations between agents and users in chronological order. We then clear the conversations data according to the following criteria. We first filter out the conversations that only contain one tweet. Furthermore, we restrict all conversations to be initialized by users, and between only one agent and one user. We also restrict that in a restored conversation the user and agent speaks in turns. In other words, user requests are followed by agents responses, and vice versa. After removing the conversations that do not meet the criteria, we have over 1.5 million conversations left, and 87.6\% of the conversations have only one round (one user request with one agent response). We then clean the conversations by removing mentions (@), hashtags (\#), and replacing URLs and numbers with ``<<url>>'' and ``<<number>>'', respectively. In our following narration, a conversation is denoted by $c$. The $i_{th}$ utterance by user is denoted by $c_i$, and its reply by agent is indicated by $a_i$. Since we restrict that a customer care conversation is initialized by a user, $c_1$ is the first user request, as well as the first utterance of the conversation. $a_1$ is the reply to $c_1$ by agent, and also the first response by agent. In this paper, we regard the utterances sent by user as ``user requests'', and the utterances by agent as ``agent responses''.

\begin{table*}
\centering
\begin{tabular}{ | l || c | c | c | c | c | c | c | c | c |  }
\hline

& $C_{Passionate}$ & $C_{Empathetic}$ & $C_{Polite}$ & $C_{Impolite}$ & $C_{Anxious}$ & $C_{Sad}$ & $C_{Satisfied}$ & $C_{Frustrated}$ \\
\hline
\hline
$R^2$  & 0.72 & 0.68 & 0.43 & 0.18 & 0.35 & 0.27 & 0.15 & 0.37\\ 
\hline
\hline

$A_{Empathetic}$ & 0.01 & 0.04 & \textbf{0.23**} & -0.01 & \textbf{-0.13***} & \textbf{-0.25**} & \textbf{0.59***} & \textbf{-0.12**}\\
\hline

$A_{Passionate}$ & \textbf{0.26***} & \textbf{0.71***} & 0.13 & 0.00 & 0.00 & \textbf{-0.15***} & \textbf{0.39**} & -0.12\\
\hline

$A_{Polite}$ & 0.02 & 0.01 & \textbf{0.49***} & 0.00 & 0.00 & 0.03 & \textbf{0.09*} & 0.03\\
\hline

$A_{Impolite}$ & 0.03 & 0.02 & -0.02 & 0.00 & 0.00 & -0.11 & -0.12 & -0.01\\
\hline

$A_{Anxious}$ & 0.02 & -0.01 & 0.07 & \textbf{0.09*} &0.02 & 0.07 & 0.04 & \textbf{0.11**}\\
\hline

$A_{Sad}$ & 0.01 & 0.00 & 0.06 & \textbf{0.04**} & \textbf{0.04**} & 0.08 & -0.09 & \textbf{0.32***}\\
\hline

$A_{Satisfied}$ & \textbf{0.38**} & 0.00 & 0.00 & -0.01 & 0.03 & 0.12 & -0.02 & 0.07\\
\hline

$A_{Frustrated}$ & 0.04 & 0.00 & 0.00 & 0.03 & \textbf{0.26***} & 0.15 & -0.02 & -0.04\\
\hline

\multicolumn{3}{l}{***$p<0.01$, **$p<0.05$, *$p<0.1$}

\end{tabular}
\caption{Results of the eight linear regression analyses on the effects of agent tones on the change of user tones. Each column represents the result of a set of regression analysis, where the dependent variable is a user tone (denoted by $C_{T^j}$), and the independent variables are eight agent tones (denoted by $A_{T^j}$). The reported results include the $R^2$ value, the regression coefficient value of each agent tone, and the significant levels of the coefficients.P values are adjusted based on Bonferroni Corrections.}~\label{tab:reg_results}
\vspace{-2em}
\end{table*}

\begin{figure}
\centering
\includegraphics[width=0.9 \columnwidth]{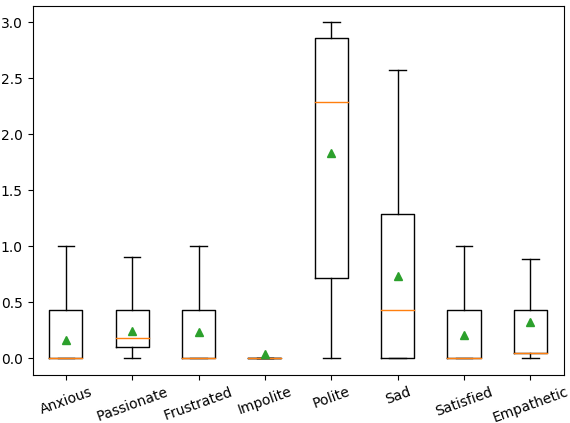}
\caption{Boxplots of the rating distributions of the eight major tones. The triangles in the plot indicate the mean values of the ratings of tones.}~\label{fig:8tones}
\vspace{-2em}
\end{figure}

\section{Formative Study}
\subsection{Major Tones in Customer Care Conversations}

To identify typical tones for customer care, we follow the processes suggested in~\cite{shabi}. We first pre-select a set of 53 tones (including aggressive, sincere, apologetic, relieved, sad, etc.) from literature across multiple domains such as marketing~\cite{aaker1997dimensions}, communication~\cite{eysenck1953structure}, linguistic~\cite{pennebaker2001linguistic}, and psychology~\cite{picard1997affective}. We analyze the intensities of the 53 tones in customer care conversations, and summarize them into eight major tones by performing factor analysis using Principal Components Analysis (PCA). According to the contributions of the 53 tones in each tone, we name the eight identified major tones as \textbf{\textit{anxious}}, \textbf{\textit{frustrated}}, \textbf{\textit{impolite}}, \textbf{\textit{passionate}}, \textbf{\textit{polite}}, \textbf{\textit{sad}}, \textbf{\textit{satisfied}}, and \textbf{\textit{empathetic}}. We then collect the intensities of these major tones in customer care conversations for studying the effects of agent tones on user experience.

Two annotation tasks are conducted for the formative study: 1) annotating intensities of the initial 53 tones for identifying major tones, and 2) annotating intensities of the identified major tones. Except for the label sets are different for the two tasks (53 initial tones vs. eight major tones), the other settings are the same. We first randomly sample 500 conversations from our dataset. We then recruit crowdworkers on CrowdFlower\footnote{https://www.crowdflower.com} to rate the intensities of tones in each utterance from all sampled conversations. The ratings are on a 4-points scale ranging from ``3: Very strongly'' to ``0: Not at all''. The tones for both tasks are arranged in random order to avoid order effect. Also, we embed validation questions in the tasks to validate annotation quality and check rater fatigue. For the workers who fail the validation questions, we filter out their answers, and ask a new worker to redo the task. Each utterance is labeled by five different valid workers, and their average rating is used as the final rating for the utterance. We restrict the workers to be native English speakers, as well as maintaining high accuracy on their previous annotation tasks. 

Figure~\ref{fig:8tones} shows the rating distributions of the major tones. Clearly, the eight major tones are commonly perceived from the customer care conversations, indicating the tones are identified properly. The most labeled tones are polite ($mean$=1.8, $std$=1.01) and sad ($mean$=0.73, $std$=0.81), indicating these two tones occur in customer care quite often. Also, passionate ($mean$=0.25, $std$=0.20) and empathetic ($mean$=0.32, $std$=0.44) tones are often perceived in the conversations. Impolite ($mean$=0.03, $std$=0.07) is the least often perceived tone in the customer care conversation, indicating that the tone is rarely occurred in customer care.

\subsection{Effects of Agent Tones}
\subsubsection{Method}
In this study, we investigate how agent tones could affect user experience. To do this, we establish eight linear regression analyses for the eight major tones. In each analysis, the dependent variable is the change of a user tone between two adjoining user requests, and the independent variables are the tones used by agents. Take user satisfied level (indicated by the satisfied tone) for an example. Let $S_{c_{i}}$ denotes the satisfied level rated by annotators in $i_{th}$ user request, i.e. $c_{i}$, in a conversation. Then the change of satisfied level between two adjoining user requests, $c_{i}$ and $c_{i+1}$, is defined as $\Delta S_{c_{i+1},c_i} = S_{c_{i+1}} - S_{c_{i}}$. The linear regression analysis on satisfied level is to study the relationship between $\Delta S_{c_{i+1},c_i}$ and the tones used by agent in $a_{i}$. Please note that, $a_{i}$ is the utterance said by agent between $c_{i}$ and $c_{i+1}$. Similarly, besides satisfied level, we also conduct regression analyses for the other major tones.

Formally, we denote a tone as $T^j$, where $j=\{1,2,\cdots, 8\}$ for eight major tones. For a tone $T^j$, we denote the rating of $T^j$ in a user request $c_i$ and a agent utterance $a_i$ as $T_{c_i}^j$ and $T_{a_i}^j$, respectively. We first compute the change of the tone between each pair of adjoining user requests for all collected conversations, denoted by $\Delta T_{c_{i+1},c_i}^j = T_{c_{i+1}}^j - T_{c_{i}}^j$. We then conduct a linear regression analysis, and the analysis takes the form:
\begin{align}
\Delta T_{c_{i+1},c_i}^j = \alpha + \beta_{1}T_{a_{i}}^{1} + \cdots + \beta_{8}T_{a_{i}}^{8}
\end{align}, 
where $T_{a_{i}}^{1}$ to $T_{a_{i}}^{8}$ are the eight selected tones used by agent in utterance $a_{i}$. $\alpha$ and $\beta s$ are the coefficients of the model. The results of the regression analyses are showed in Table~\ref{tab:reg_results}. For a certain dependent variable (i.e. change of a user tone), the table lists three types of information: 1) $R^2$, indicating how well the dependent variable can be explained by the independent variables (i.e. agent tones), 2) \textit{regression coefficients}, measuring the effects of the independent variables on the dependent variable, and 3) \textit{P value}, indicating if the effects are statistically significant. We applied Bonferroni correction on the significance levels to reduce the risk of Type II errors.

\subsubsection{Results}

Overall, all eight linear regression analyses report reasonable $R^2$ values, indicating the strong relations between the changes of user tones and agent tones. The regression analysis on the change of user tone \textbf{\textit{passionate}} and \textbf{\textit{empathetic}} reports highest $R^2$ value of 0.72 and 0.68, respectively. It indicates that the changes of these two user tones can be well explained by the tones used by agent. The analyses on another four tones -- \textbf{\textit{anxious}}, \textbf{\textit{polite}}, \textbf{\textit{sad}}, and \textbf{\textit{frustrated}} report reasonable $R^2$ values of 0.35, 0.43, 0.28, and 0.37, respectively. The $R^2$ value of \textbf{\textit{satisfied}} is 0.15, indicating that a moderate proportion of variability of user satisfaction can be explained by agent tones used in the responses. We suggest that this is because user satisfaction may more depend on whether their issues are resolved, hence is less affected by how agents talk. Similarly, the changes of \textbf{\textit{impolite}} tone of users are also moderately explained by the linear model, with a $R^2$ value of 0.18. This may be because people being impolite in their language is mainly due to their characters and habits, and has less to do with how they are responded. Next, we report the effect of each agent tone on user experience as follow.

\textbf{\textit{Empathetic}}: Our results indicate that agent using empathetic tone has significant effects on the changes of five user tones. First, it increases positive emotion of users, such as satisfaction ($coef=0.59, p<0.01$) and politeness ($coef=0.23, p<0.05$) in user requests. Meanwhile, it reduces negative emotion of users, including anxiety ($coef=-0.13, p<0.01$), sadness ($coef=-0.25, p<0.05$ ), and frustration ($coef=-0.12, p<0.05$). 

\textbf{\textit{Passionate}}: The regression analysis also reports the benefit of applying passionate tone in customer care. It significantly affects the changes of four user tones. The tone has significant positive effects on the changes of satisfied ($coef=0.39, p<0.05$), passionate ($coef=0.26, p<0.01$), and empathetic ($coef=0.71, p<0.01$) tone in user requests. Also, it significantly reduces user sadness ($coef=-0.15, p<0.01$).

\textbf{\textit{Polite}}: Polite tone only marginally significantly increases user satisfaction ($coef=0.09, p<0.1$), and the effect is much lighter than former two tones, i.e. the regression coefficient is much smaller. This implies agent being polite only has limited effect on user satisfaction level. Meanwhile, agent being polite could cause user to be polite too ($coef=0.49, p<0.01$). The results also indicate that agents using polite tone does not have significant effect on reducing any negative tones of users. 

\textbf{\textit{Satisfied}}: The satisfied tone used by agents significantly increases the passionate tone in user request ($coef=0.38, p<0.05$). The tone does not have significant effect on the other user tones. 

\textbf{\textit{Impolite}}: Impolite tone does not show significant effect on any user emotion. We suppose this is because agents rarely appear to be impolite during customer care (Figure~\ref{fig:8tones}). 

\textbf{\textit{Anxious}}: The results indicate that agents showing anxious in their responses makes users feel more frustrated ($coef=0.11, p<0.05$), and also marginally significantly increase users' impolite tone ($coef=0.09, p<0.1$). There is no other significant effect of anxious tone is observed in the results. 

\textbf{\textit{Frustrated}}: Frustrated tone of agents is also reported of negative effects for customer care -- it significantly increases the anxiety level of users ($coef=0.26, p<0.01$). There is no other significant effect of this agent tone observed.

\textbf{\textit{Sad}}: Sad tone in agent responses significantly increases user frustration ($coef=0.32,p<0.01$), and slightly increases user impolite ($coef=0.04, p<0.05$), and anxious ($coef=0.04, p<0.05$) tones. The tone does not have significant effect on other user tones. 

Overall, the linear regression analyses indicate that applying positive tones, especially \textbf{\textit{empathetic}} and \textbf{\textit{passionate}} tones, are beneficial for customer care. Applying these tones in responses could significantly reduce user negative emotions such as anxiety, frustration, and sadness, and increase user satisfaction. 
Therefore, we target to embedding \textbf{\textit{empathetic}} and \textbf{\textit{passionate}} tones in the responses of our tone-aware chatbot.

\begin{table}
\centering
\begin{tabular}{  l || c  }
\hline

\textbf{\textit{Empathetic}} & \textbf{\textit{Passionate}} \\
\hline
sorry*** & !***\\
\hline
apologize*** & :)***\\
\hline
frustration*** & great**\\
\hline
confusion*** & glad**\\
\hline
inconvenience** & wonderful**\\
\hline
hear** & directly**\\
\hline
understand** & certainly**\\
\hline
concerns** & hey**\\
\hline
happened** & definitely**\\
\hline
patience** & sure**\\
\hline
details** & exactly**\\
\hline
aware** & much**\\
\hline
\multicolumn{2}{l}{***$p<0.01$, **$p<0.05$, *$p<0.1$}

\end{tabular}
\caption{Example keywords for empathetic and passionate tones, as well as the significant level from t-test. P values are adjusted based on Bonferroni corrections.}~\label{tab:key-word}
\vspace{-2em}
\end{table}

\subsection{Tone Keywords}
To endow our chabot with capacity of responding with tones, further information on how these beneficial tones are expressed in customer care conversations is required. Therefore, in this section, we extract the keywords of the two selected tones. To do this, we first collect all commonly used uni-grams, bi-grams, and tri-grams from all agent responses that used more than ten times. The frequency of each term in an utterance is then computed. Specifically, the frequency is calculated as the number of occurrence of a term in a utterance divided by the number of all terms in the utterance. We next find all empathetic and passionate agent responses according to their ratings from the human labeled dataset. Since 3-points indicate the intensity of ``moderately strong'' in our 4-points scale, we take 3-points as a threshold. According to the empathetic scores, we split agent responses into two datasets -- empathetic and non-empathetic datasets. Similarly, we split agent responses into passionate and non-passionate datasets according to their scores on passionate tone. Taking empathetic for example, we aim to finding the keywords that occur in empathetic responses and non-empathetic responses with significantly different frequencies. Therefore, we conduct a t-test for each term between its occurrence frequencies in empathetic responses and the frequencies in non-empathetic responses. We take the terms that pass the tests at a significant level of 0.05 as empathetic keywords. We use the same method to locate the keywords for passionate tone. We list example keywords for both tones in Table~\ref{tab:key-word}, along with their significant level from t-tests adjusted by Bonferroni corrections.

Overall, all the extracted keywords are uni-gram for both empathetic and passionate tones. We suggest that this is because the dataset is relatively small, therefore, bi-grams and tri-grams do not occur frequently in the data. We find 28 keywords for empathetic tones. Among these words, we mainly observe two word types. The words of the first type are used by agents for apologizing, such as ``sorry'', ``apologize'', ``really'', and ``happended''. The second type consists of the words indicating agents' understanding to users' suffering, for example ``understand'', ``inconvenience'', ``frustration'', and ``confusion''. There are 19 words identified as passionate keywords. There are also two types of words observed. The first type of keywords such as ``!'', ``definitely'', ``certainly'' is used for increasing the certainty and trustworthiness in agent responses. The second type of words are positive emotional words such as ``great'', ``awesome'', and ``love''. 

\begin{figure*}
\centering
\includegraphics[width=1 \textwidth]{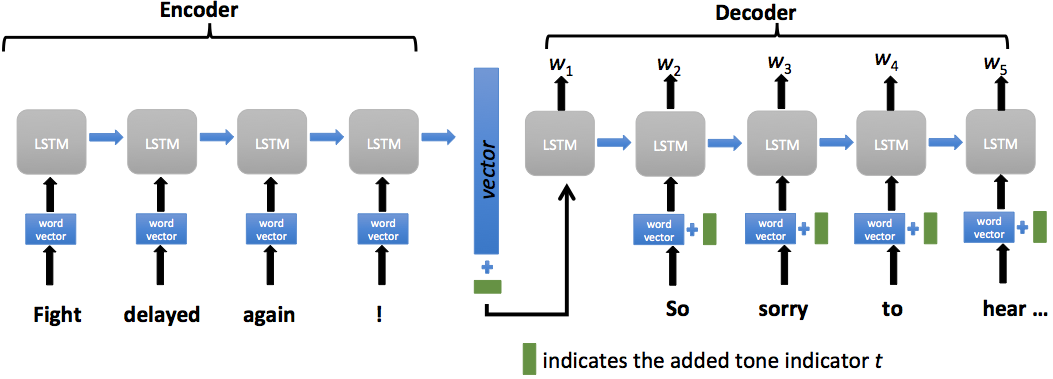}
\caption{An illustration of the structure of our tone-aware seq2seq model. The figure shows the training process, in which the encoder and decoder take a user request and an agent response as the inputs, respectively. Meanwhile, a tone indicator $t$ is deployed for training the decoder to learn the expression patterns for different tones.}~\label{fig:seq2seq}
\vspace{-2em}
\end{figure*}

The usage of extracted keywords in customer care conversations implies how passionate and empathetic tones are expressed by human agents. A tone-aware chatbot should be able to learn such information, and apply it on response generating. To effectively embed the information into our chatbot, we deploy a novel seq2seq model. The model takes the keywords as indicators of different tones, and learns how to generate response of similar styles.

\section{Tone-aware Chatbot}

Deep learning based techniques, such as seq2seq models, are widely adopted for automatic conversation generation~\cite{sutskever2014sequence}. In general, a typical seq2seq model is trained on data of sequence pairs. From training process, the model learns the matching relation between sequence pairs, and gains the ability of generating a matching sequence for a given sequence. In terms of customer care chatbot, the training is based on customer care conversation data, and chatbot learns how to respond to given user requests. A standard seq2seq model cannot handle meta information of data in the learning process. Consequently, the model cannot offer a better control on the generated results, e.g. control the tones of generated responses. To integrate tone information in our chatbot, we propose a novel seq2seq model in this work. The model takes tone information of conversations as a type of input, and learns the various expressions of different tones. Therefore, given an assigned tone, the chatbot is able to generated responses in the tone.


\subsection{Background on Seq2seq Learning}

A standard seq2seq model consists of two parts: a encoder and a decoder. Both encoder and decoder are implemented with recurrent neural network (RNN), such as LSTM model~\cite{sundermeyer2012lstm} or GRU model~\cite{chung2014empirical}. Since RNN is able to keep track of historical information, the structure particularly suits for dealing with sequential data. We select LSTM in this work. 

In the training process, the input of the model are two matching sequences, denoted by $X = \{x_1, x_2, \cdots, x_n\}$ and $Y = \{y_1, y_2, \cdots, y_m\}$. For example, $X$ is a user request and $Y$ is the agent response, where $x_i$ and $y_i$ are words in the two utterances. The encoder and decoder are trained on $X$ and $Y$ respectively. The connection between these two sequences is established by passing a vector representation of $X$ from the encoder to the decoder. Specifically, the first sequence $X$ is embedded to a vector $h$ by the encoder. Both $h$ and the second sequence $Y$ are then fed to the decoder. By doing this, the decoder learns the sequential patterns of $Y$ given the contextual information from $X$. In the generating process, the input of the model is only the first sequence $X$. Similar to the training, $h$, a vector representation of $X$, is computed by the encoder. The decoder takes the vector $h$ as its only input. At each step, the decoder generates a new element $w_i$. The generating process keeps going until it reaches the maximum step count or an end indicator. The generated sequence $W = \{w_1, w_2, \cdots, w_m\}$ is the output of the model.

\begin{table*}[!ht]
\centering
\begin{tabular}{ | l || l | }
\hline
\textit{\textbf{Request}} & By far the best mag cover! Their flight attendants are always funny friendly helpful flysouthwest <<url>>\\
\hline
\textit{\textcolor{YellowOrange}{\textbf{Passionate}}} & \textcolor{YellowOrange}{We love hearing that! We'll be sure to pass along your kudos to our crew. Thanks for flying with us!} \\
\hline
\textcolor{OliveGreen}{\textit{\textbf{Empathetic}}} & \textcolor{OliveGreen}{We are happy to hear you like the service you deserve. Please let us know if you have any further concerns.} \\
\hline
\textit{\textcolor{gray}{\textbf{Neutral}}} & \textcolor{gray}{Thanks for the shout out.} \\
\hline
\textit{\textbf{Human}} & We're glad you like it holly! We have some amazing employees out there. \\
\hline
\hline
\textit{\textbf{Request}} & Do you charge overseas transaction fees? If so what is the charge? \\
\hline
\textit{\textcolor{YellowOrange}{\textbf{Passionate}}} & \textcolor{YellowOrange}{Great question! You can view the transaction on the account tab in the pnc olbanking.} \\
\hline
\textit{\textcolor{OliveGreen}{\textbf{Empathetic}}} & \textcolor{OliveGreen}{Hi what happened? Anything I can do to help?} \\
\hline
\textit{\textcolor{gray}{\textbf{Neutral}}} & \textcolor{gray}{Hi there. For assistance with your inquiry please contact our credit card team at <<number>> for assistance.} \\
\hline
\textit{\textbf{Human}} &  Hi. Which card product do you have? Pls do not release any personally identifiable information.\\
\hline
\hline
\textit{\textbf{Request}} & Really <number> min warning 4th ur shit has to go out. Direct tv here I come \\
\hline
\textit{\textcolor{YellowOrange}{\textbf{Passionate}}} & \textcolor{YellowOrange}{Oh no! We don't want you to feel this way! Please dm* us your acct so we can look into your service issues.}\\
\hline
\textit{\textcolor{OliveGreen}{\textbf{Empathetic}}} & \textcolor{OliveGreen}{I would be happy to address any service concerns. Dm me your account number or full address so I can help.}\\
\hline
\textit{\textcolor{gray}{\textbf{Neutral}}} & \textcolor{gray}{Can you please send me dm with your acct or phone so I can assist you?}\\
\hline
\textit{\textbf{Human}} & Hey there! Is there anything we can do for you today? Dm us!\\
\hline
\hline
\textit{\textbf{Request}} & Your website is down and I really need to buy chair! ! ! No joke.\\
\hline
\textit{\textcolor{YellowOrange}{\textbf{Passionate}}} & \textcolor{YellowOrange}{Hi there! We are experiencing technical issues with our website. We are working to resolve this.}\\
\hline
\textit{\textcolor{OliveGreen}{\textbf{Empathetic}}} & \textcolor{OliveGreen}{Hi I apologize for the inconvenience. Can you please dm your account or phone number?}\\
\hline
\textit{\textcolor{gray}{\textbf{Neutral}}} & \textcolor{gray}{Hi there. Can you dm us your number so we can look into this for you?}\\
\hline
\textit{\textbf{Human}} & Hi Inga we were aware that the website was down and we apologize for this. It is now up and running again.\\
\hline
\hline
\textit{\textbf{Request}} & Thanks to for the twitter triage. Shouts to reps dsc and qs for the one-on-one help.\\
\hline
\textit{\textcolor{YellowOrange}{\textbf{Passionate}}} & \textcolor{YellowOrange}{You're welcome!}\\
\hline
\textit{\textcolor{OliveGreen}{\textbf{Empathetic}}} & \textcolor{OliveGreen}{We are glad to hear that you were able to get the help you needed.}\\
\hline
\textit{\textcolor{gray}{\textbf{Neutral}}} & \textcolor{gray}{Hi I am happy to help with the billing. Can you please send me DM with your acct or phone?}\\
\hline
\textit{\textbf{Human}} & I'm glad me and qs could help out today! =) \\
\hline
\hline
                 & \textit{User}: my WiFi not working\\
\textit{\textbf{Request}} & \textit{Agent}: We'd love to help you! Where are you located? We can check for outages in the area.\\
      & \textit{User}: Orlando Florida\\
\hline
\textit{\textcolor{YellowOrange}{\textbf{Passionate}}} & \textcolor{YellowOrange}{Thanks for getting back to us! We have responded to your DM. Please check your DM.}\\
\hline
\textit{\textcolor{OliveGreen}{\textbf{Empathetic}}} & \textcolor{OliveGreen}{Thank you for your patience. We are working to restore service quickly.}\\
\hline
\textit{\textcolor{gray}{\textbf{Neutral}}} & \textcolor{gray}{If you can DM me your order number I will look into this for you.} \\
\hline
\textit{\textbf{Human}} & So that we can accurately assist you, do you have DSL or U-Verse internet?\\ 
\hline
\hline
                 & \textit{User}: Can you help me get my money back that you guys are keeping hostage\\
\textit{\textbf{Request}} & \textit{Agent}: I am so sorry to hear that t-mobile is has not given you your monies back. Is it for over payment?\\
      & \textit{User}: I ordered something, My card was not supposed to be charged until device shipped. \\
\hline
\textit{\textcolor{YellowOrange}{\textbf{Passionate}}} & \textcolor{YellowOrange}{I'm sorry to hear that! If you haven't already please reach out to us here <<url>>}\\
\hline
\textit{\textcolor{OliveGreen}{\textbf{Empathetic}}} & \textcolor{OliveGreen}{I understand your frustration. If you would like to cancel the order please contact us here <<url>>}\\
\hline
\textit{\textcolor{gray}{\textbf{Neutral}}} & \textcolor{gray}{Thank you for the information. Please allow us to research this for you .} \\
\hline
\textit{\textbf{Human}} & I totally understand your needing your money asap, was it from an iPhone order?\\ 
\hline
\hline
\end{tabular}
\caption{Examples of the agent responses of different tones to real-word user requests generated by our system. We list the passionate responses in \textcolor{YellowOrange}{orange}, empathetic responses in \textcolor{OliveGreen}{green}, and neutral responses in \textcolor{gray}{gray}. We also list responses by real human agents in black for comparison. The table shows seven conversations, five of them are of one round, and the other two are of multiple rounds. Please note that to ensure the evaluation is based on real customer care data, we only generate responses for the last round for multiple rounds conversation. By doing this, both human agents and the chatbot are given the same contexts (previous rounds), hence we are able to fairly evaluate our chatbot by comparing the generated responses with human responses. *``DM'' in the conversations indicates ``Direct Message'' in Twitter.}~\label{tab:examples}
\vspace{-2em}
\end{table*}

\subsection{Tone-aware Seq2seq Learning}

To enable our chatbot to respond in different tones, the underlying model is designed not only to learn how to respond user requests, but also to learn the different expressions of tones. Therefore, besides the request and response pairs, the tone information is also required for training the model for differentiating the tones. Assume that a request \textit{``My fight is delayed again!''} and its response \textit{``I know delays are frustrating.''}, as well as the tone information (empathetic) are fed into our model. Similar to a standard model, the model first learns how the responses are constructed according to the requests (regular response patterns). More importantly, it learns the different language patterns associated with different tones. For example, the model may discover that when the tone is empathetic, the response are more likely to use ``frustrating''. The model emerges the learned tone expression patterns with the regular response patterns, and eventually results in the ability of respond in different tones. Please note that, the model does not simply insert the representative words into responses based on fixed rules, but learns how to organically use tone expressions in responses. Therefore, when the trained chatbot is asked to reposed to a request, for example \textit{``I've been waiting for my package for weeks!''}, the model could combined the learned patterns, and generate a empathetic response, such as \textit{``We know waiting is frustrating.''}

Our model is inspired by~\cite{li2016deep}, in which the decoder is modified to handle meta information by add an embedding vector. In this work, we adopt the idea, except we use a simple indicator to code the meta information instead of a embedded vector. Let $X = \{x_1, x_2, \cdots, x_n\}$ and $Y = \{y_1, y_2, \cdots, y_m\}$ again denote two matching sequences. 
$t$ is an indicator vector of length one recording the meta information of the sequence pair. In this work, our chatbot models two selected tones -- empathetic and passionate, and we also model the conversations that are neither (neutral tone). Therefore, $t$, in our case, takes three values to indicate the tones. The real values of $t$ does not affect our model, as long as different values are assigned for different sequence types.

For the training process, the encoder acts the same as a standard seq2seq model. The first sequence $X$ is encoded to a vector presentation $h$. In the standard model, the input sequence for the decoder is $\{h, v(y_1), v(y_2), \cdots, v(y_m)\}$, where $v(y_i)$ is the embedding vector for $y_i$. In our tone-aware model, we concatenate $t$ to each vector of the input sequence for the decoder. The new input sequence is denoted as $\{h', v(y_1)', v(y_2)', \cdots, v(y_m)'\}$, where $h' = h \oplus t$, and $v(y_i)' = v(y_i) \oplus t$. Note that $\oplus$ indicates vector concatenate operation. In other words, the meta information is added to each step for the decoder in the training process. By doing this, the decoder not only learns the sequential patterns of $Y$, but also keeps track of the differences between different types of input sequence. 

In the generating process, the input includes $X$ and $t$. Note that, $t$ in generation indicates the type we want for the output sequence, i.e. a specific tone for an agent response. The encoder takes $X$ as the input and outputs a vector $h$. Then $h$ is concatenated with $t$, and the results $h' = h \oplus t$ is fed into the decoder. For each step, the decoder generates an element $w_i$. The embedding vector of the generated element $v(w_i)$ is also concatenated with $t$, and the result $v(w_i)' = v(w_i) \oplus t$ is fed into the decoder to generate the next element $w_{i+1}$. Similarly to a standard seq2seq model, the generating process stops at the maximum step count or an end indicator, and outputs the generated sequence $W = \{w_1, w_2, \cdots, w_m\}$. Figure~\ref{fig:seq2seq} illustrates the tone-aware seq2seq model introduced in the paper.

\begin{figure*}
\centering
\includegraphics[width=1 \textwidth]{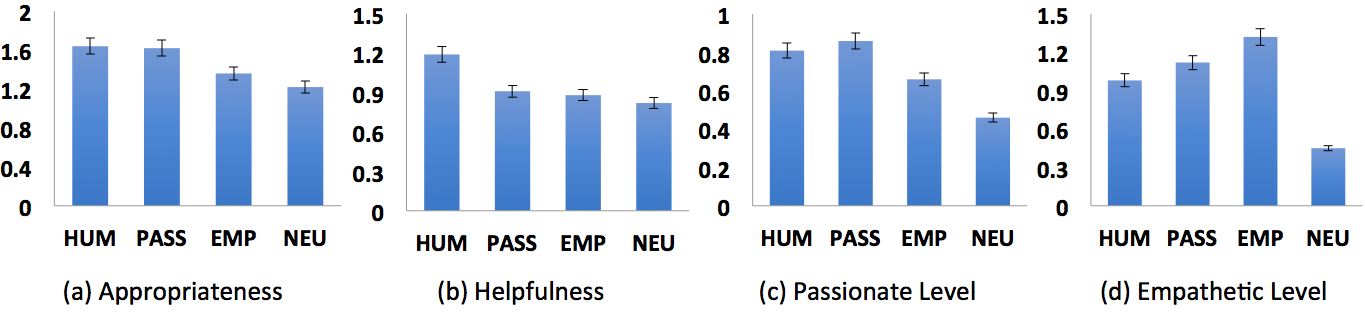}
\caption{Results of human evaluation. Please note that, ``HUM'' is short for ``Human'', indicates the real human agent response, ``PASS'' is short for ``Passionate response'', ``EMP'' is short for ``Empathetic responses'', and ``NEU'' represents ``Neutral responses''.}~\label{fig:user_study}
\vspace{-2em}
\end{figure*}

\subsection{Implementation}

We first preprocess the conversations data into the form of sequence pairs with tone information for both training and generating. For the conversations of one round (one user request, $c_1$, with one agent response, $a_1$), the sequence pairs are simply these two utterances, where $X = c_1$ and $Y = a_1$. For the conversations of multiple turns, we generate a pair of matching sequences for each round. The first sequence $X$ for $i_{th}$ round is defined as the user request in this round $c_i$, plus all the utterances by both the user and agent before the round. Formally, $X = c_1 \oplus a_1 \oplus c_2 \oplus a_2, \cdots, \oplus c_i$. The second sequence $Y$ is simply the agent response in this round, i.e. $Y = a_i$. This method of generating sequence pairs from multiple turns conversations is to include the context for each turn as suggested in~\cite{li2016deep}.

The extracted keywords are used for indicating the conversation tone information. Specifically, for a preprocessed conversation, if one or more empathetic word occur in the agent response ($a_1$ for one round or $a_i$ for multiple rounds), $t$ is set to an empathetic indicator. Similarly, $t$ is set to a passionate indicator, if passionate keywords occur. If no keyword occurs in the agent response, implying the conversation does not contain tone information, $t$ is then set to a neutral indication.


From the preprocessed sequence pairs, we randomly sample 500 pairs for evaluation, and the remaining of the data is used for training. For an evaluation sequence pair, we fed the user request to the model three times, each time with a different value for the tone indicator $t$. In other words, for each evaluation user request, we generate a response in empathetic, passionate and neutral tone, respectively. Table~\ref{tab:examples} shows some evaluation user requests, and the generated responses by our chatbot, as well as the responses by real human agents.

The configurations of the models are as follow:

1) One layer LSTM models for both encoder and decoder with 512 hidden cells. Parameters of the model are initialized by sampling from the uniform distribution [-0.1, 0.1].

2) The vocabulary size is set to 10,000, and the size of word embedding is 256. The tone indicator $t$ is set to -1, 0, +1 for empathetic, neutral, and passionate tones, respectively. 

3) The model is trained using Adam optimization algorithm~\cite{kingma2014adam} with an initial learning rate of 0.001.



\section{Evaluation}

We evaluate our system from two aspects: 1) the response quality, i.e. if the system generates proper responses to user requests, and 2) the intensities of embedded tones, i.e. if human could perceive tones embedded in the generated responses. To assess the response quality, we derive two metrics from previous work: response \textbf{\textit{appropriateness}} and \textbf{\textit{helpfulness}}. Specifically, an appropriate response should be on the same topic as the request, and should also ``make sense'' in response to it~\cite{ritter2011data}. A helpful response should contain useful and concrete advice that can address the user request~\cite{harper2008predictors}. To assess the intensities of embedded tones, we evaluate \textbf{\textit{passionate level}} and \textbf{\textit{empathetic level}} for the generated response. 

Given a user request, three responses are generated by our system for three different tones (passionate, empathetic, and neutral). We then ask annotators to rate the appropriateness, helpfulness, passionate level, and empathetic level for the three generated responses, as well as the real agent response to the same user request. By comparing the ratings of generated responses with real agent responses, we validate to what extent our system could generate human-like responses to user requests, in terms of both response quality and tone intensity. Meanwhile, by comparing the ratings of three generated responses, we validate if our system is capable of embedding different tones in the responses.



Crowdflower is again used to recruit annotators. All 703 participants are native English speakers, and they are 18 or older. 
Participants are asked to fill out at least one gold question in order to participate the survey. 14.1\% of participants fail the check, and their responses are removed. In a survey task, participants are first instructed to learn the four rating metrics: appropriateness, helpfulness, passionate and empathetic levels, with definitions and examples. Then, they are shown a user request and asked to rate the four responses. The responses are arranged in random order to control order effects. 500 user requests are sampled, thus 2,000 responses in total are rated. Each response is rated by five participants according to the four metrics. The ratings are made on a 7-point scale from strongly disagree (-3) to strongly agree (+3). The average of five participants' ratings of a response are used as the final ratings. Intra-class correlation (ICC) is used to assess the reliability for our evaluation tasks, where each response is rated by $k$ workers randomly selected from a population of K workers. ICC(1,$k$) is ranged from 0.56 to 0.87, indicating moderately high reliabilities of the workers~\cite{biel2011you}.

\subsection{Results}

For each of the four criteria, we first conduct an ANOVA analysis, respectively. All four ANOVA analyses reject the null hypothesis, indicating that scores on different types
of responses (passionate, empathetic, neutral, real human agent response) are significantly different. Therefore, additional post analysis is then conducted for each metric. To be specific, in the post tests, we compare the different types in pairs using t-test for each criterion, and apply Bonferroni correction on the results. We report the post analysis results as follow.


\textbf{\textit{Appropriateness}}: Interestingly, there is no statistically significant difference between human responses and generated passionate responses in terms of appropriateness ($p=0.33$). This indicates that our chatbot has a similar ability as human agents to respond appropriately to requests. However, the appropriateness scores of three types of toned response are significantly different. Passionate responses are rated significantly higher than empathetic ($p<0.05$) and neutral ($p<0.05$) responses, indicating that passionate tone is considered more appropriate to address requests by the annotators. We suggest that this is because passionate tones can address more situations of user requests than the other two tones. We will discuss the difference among tones in the next section. Meanwhile, it is observed that empathetic responses are marginally significantly more appropriate than neutral responses ($p<0.1$). 

\textbf{\textit{Helpfulness}}: Not surprisingly, real human responses are rated higher than generated empathetic ($p<0.01$), passionate ($p<0.01$), and neutral ($p<0.01$) responses. This is because helpfulness measures if the responses contains useful and concrete advice. In other words, to achieve a higher helpfulness score requires more background knowledge and extra information~\cite{xu2017new}. It is reasonable that human agents do a better job than our system, which is trained on a finite dataset. There is no significant difference reported among the three generated responses. It implies that using different tones in a response do not affect the helpful level.

\textbf{\textit{Passionate Level}}: There is no statistically significant difference observed between human responses and generated passionate responses in terms of passionate level ($p=0.15$). The results indicate that our chatbot can perform as passionate as human agents do. As for the three tone types, empathetic responses are lower rated than passionate responses ($p<0.01$), and neutral responses' rates are lower than empathetic responses ($p<0.01$). The results imply that our chatbot successfully controls the passionate tone in the generated responses, and the human annotators are able to perceive the tone.

\textbf{\textit{Empathetic Level}}: Surprisingly, annotators perceive that our system generates the most empathetic responses. All three t-tests reveal that empathetic responses are significantly higher rated comparing with the other response types ($p<0.01$ for all three tests). Passionate responses are rated the second in terms of empathetic level, and marginally significantly higher than human response ($p<0.1$), and significantly higher than neutral responses ($p<0.01$). Human response is rated the third, and perceived more empathetic than neutral responses ($p<0.01$). This indicates that our system also controls empathetic tone well, and could even perform more empathetically than human agents do.

In summary, the test results reveal that our system could perform as well as human agent in term of responding to users properly. However, the annotators rate our chatbot less helpful than real human agents. The tones embedded in the responses by our chatbot can be accurately perceived by human. The intensities of the embedded tones in the generated responses are as strong as (passionate), or even stronger (empathetic) than the intensities of tones in human responses.

\section{Discussion}

In this work, we report our tone-aware chatbot for social media customer care. According to the human evaluation results, our chatbot is able to embed human-perceivable tones in responses. Meanwhile, the tone-aware chatbot performs reasonably in terms of the appropriateness and helpfulness levels. 
Since tones used by agent in their responses could significantly affect user experience, our work is of much practical value. A possible application of our system is recommending human agents responses in different tones to a user request. By doing this, the agent reacting time could be significantly reduced, and tones are also considered by the system.

We summarize and identify eight major tones that commonly occur in customer care conversation. Our study on the eight major tones suggests that two tones -- empathetic and passionate -- are of great value to customer care. The benefits of using these tones include reducing user negative emotion, increasing their positive emotion, and eventually increasing user satisfaction. Not surprisingly, the tones embedded by our tone-aware chatbot appear to have similar effects of increasing user experience. According to the human evaluation results, both empathetic and passionate responses are scored significantly more appropriate than neutral responses. The observation implies the effects of tones, and more importantly, evidences the application value of the tone-aware chatbot.

Statistically, the evaluation indicates that the passionate responses generated by our chatbot are perceived as appropriate as the responses by human agents. Interestingly, according to previous work on customer care chatbot~\cite{xu2017new}, to a certain extent, human responses still outperform generated responses on appropriateness. We suppose this is probably due to the passionate tone embedded in the responses. According to our formative study, passionate tone has significant positive effect on user experience. Therefore, we suggest that the passionate tone embedded by our chatbot may increase the impressions of the responses to the annotators, and consequently lead to higher appropriateness scores. 

The results of evaluation also reveal the differences between two tones controlled by the system. Passionate tone is rated significantly more appropriate than empathetic tone. We suggest that this is because passionate tone is more general. In other words, this tone could fits for more types of user requests. For example, Xu et al. suggested in~\cite{xu2017new} that user request could be informational (seeking for information), or emotional (expressing emotion). For informational user request, it may be not that appropriate for agent to show empathetic tones, while passionate tone could be still appropriate to use. Hence, passionate tone could be assigned higher rating than empathetic tone in these cases. It is worth studying the effects of different tones at a finer granularity in the future work, and furthermore, how to adopt the fine-grained effects in an automatic conversational customer care system.

Our evaluation is based on human judgment, instead of automatic metrics. This is because although automatic metrics can be used to evaluate if the generated results are grammatically correct, they are not capable of evaluating the embedded tones. Therefore, human judgment is required for evaluating if users can perceive the tones we embed in the responses. Moreover, the grammatical correctness is also evaluated by comparing the appropriateness and helpfulness of the generated responses and the real agent responses. In the future, additional studies can be designed to examine how our chatbot is used in practice. For example, a field study can be designed to study the perceptions of perceived tones from end users of a brand, and how these perceptions can affect user engagement.

The proposed tone-aware chatbot is based on a deep learning seq2seq framework. The model controls the output styles by simply adding an indicator bit to the standard seq2seq model. Besides tones, the model can be easily extended to generate responses of various styles. A possible and interesting extension is a brand-aware chatbot for customer care. The chatbot reported in this work answers user request in a general brand style. However, customer care of different brands may have different styles. For example, according to our observation on the collected dataset, the customer care of some brands such as Nike (@Nike) and Jetblue (@JetBlue) tend to use informal responses, while some others are more formal, such as Chase (@Chase) and Wells Fargo (@WellsFargo). Therefore, it is possible to extend our model to a brand-aware chatbot by feeding the model meta information of brands. The chatbot could learn the styles of different brands accordingly, and generates agent responses in different brand styles.  

\section{Conclusion}
In this paper, we first systematically study the effects of agent tones in customer care. Two tones that are beneficial for increasing user experience -- passionate and empathetic -- are identified accordingly. We further propose a novel deep learning based chatbot for customer care that integrates the tone information in conversations, and generates toned responses to user requests. The evaluation results suggest that our system could generate as appropriate responses as human agents. Meanwhile, the the tones embedded can be easily perceived by annotators. More importantly, it is observed that the responses generated by our system is perceived more empathetic than responses by human agents. There are many interesting and valuable directions for future work. Possible directions include studying the effects of agent tones at a finer granularity, and how the chatbot could effect the end user engagement. Meanwhile, it is also worth studying the possible extensions on our proposed model, such as a brand-aware chatbot for different brand styles.


\bibliographystyle{SIGCHI-Reference-Format}
\bibliography{sample}

\end{document}